\documentclass[runningheads]{llncs}
\usepackage[T1]{fontenc}
\usepackage{adpwrapfig}
\usepackage{graphicx}
\usepackage{algorithm2e}
\usepackage{amsmath}
\usepackage{csquotes}
\usepackage{cite}
\usepackage{setspace}
\usepackage[binary-units=true]{siunitx}
\usepackage{tabularx}
\usepackage{caption}

\usepackage[hidelinks]{hyperref}

\usepackage{color}

\SetAlCapSkip{1em}
\allowdisplaybreaks

\DeclareSIUnit{\nothing}{\relax}

\DeclareMathOperator{\data}{data}
\newcommand{\serialrate}{r_{s}}
\newcommand{\parallelfactor}{r_p}
\newcommand{\parallelizability}{p}

\begin{document}
\title{Modeling Task Mapping for Data-intensive Applications in Heterogeneous Systems}
\titlerunning{Modeling Task Mapping for Data-intensive Appl. in Heterogeneous Systems}
%
\author{Martin Wilhelm\inst{1} \and
Hanna Geppert\inst{1} \and
Anna Drewes\inst{1} \and
Thilo Pionteck\inst{1}}
\authorrunning{M. Wilhelm et al.}
%
\institute{Otto-von-Guericke University, Magdeburg, Germany}
\maketitle              

\begin{abstract}
We introduce a new model for the task mapping problem to aid in the systematic design of algorithms for heterogeneous systems including, but not limited to, CPUs, GPUs and FPGAs. A special focus is set on the communication between the devices, its influence on parallel execution, as well as on device-specific differences regarding parallelizability and streamability.
We show how this model can be utilized in different system design phases and present two novel mixed-integer linear programs to demonstrate the usage of the model.

\keywords{heterogeneous computing, task mapping, resource allocation, modeling, MILP, FPGA, hardware/software partitioning, design space exploration}
\end{abstract}

\section{Introduction}

With Moore's Law declining, modern computing systems become increasingly heterogeneous, containing processing devices, such as CPUs, GPUs or FPGAs as well as associated memories with vastly different characteristics. A significant challenge lies in the mapping of application tasks to fitting devices. In general, a mapping should minimize the execution time of a task on a certain device, which is influenced by multiple factors, such as the parallelizability and streamability of a task. Nevertheless, a better suited device may be a suboptimal choice if the device is already highly contended. Moreover, even an unused device with a high processing speed may be avoided if the data transfer cost between devices exceeds the gain from the parallelization. 

In this work, we develop an abstract model for the task mapping problem on heterogeneous devices for data-intensive applications, where communication cost plays a significant role. With this model we aim to support developers in the early design phase of a heterogeneous system and clear the path for theoretical evaluations and comparisons of task mapping algorithms. We demonstrate the capabilities of the model based on two linear programs in a sample environment, which can be used as a reference for future heuristics.

\section{State of the art}

The mapping of tasks to processing devices (also called resource/task allocation or workload partitioning) describes a central step in the design of heterogeneous systems. Much work exists for CPU-GPU task mapping~\cite{mittal2015}. Research in this field mainly focusses on (application-)specific algorithms without a reference to a general model or a common measure of cost. This makes it difficult to compare different approaches and to transfer insights to new problems. Some authors introduce a more detailed model~\cite{campeanu2014,wang2020}. However, the underlying parallelism of a heterogeneous system is seldom taken into account, especially with respect to the impact of data transfer. In the field of production research, a closely related problem is known as the \emph{agent bottleneck generalized assignment problem}~\cite{karsu2012, bektur2019}. Here, the parallel execution through different agents is central, but communication cost between the agents are usually not present.

Few work is present that includes dataflow-based devices such as FPGAs. Works that include FPGAs frequently model them similar to software processing units~\cite{Wang2021}. Yet, FPGAs have special characteristics as they are area-bound and enable pipelining, leading to vastly different behavior. Modeling these differences is crucial for exploiting their full potential~\cite{che2008}. Owaida et al. discuss these differences in the context of designing OpenCL tasks for FPGAs~\cite{owaida2015}. Much work is done in the closely related field of hardware/software partitioning~\cite{mhadhbi2016}. Models in this field better reflect hardware properties~\cite{niemann1997}, but usually do not differentiate between software units e.g.\ in terms of parallelizability.

\section{Modeling}

In this section, we develop an abstract system model with a minimal set of interfaces that allows us to define a cost function to assess the quality of a given task mapping. We then show how this model can be utilized in different phases of a systematic design space exploration for a heterogeneous system. 

\subsection{Abstract model}
\label{ssc:abstractmodel}

In different design phases, different knowledge about the system properties is present, therefore it is crucial to make single components of the model exchangeable without the need to adjust other components or the underlying algorithm. For this, we split the system model into an \emph{application model}, which describes the properties of and relations between tasks, a \emph{platform model} describing the characteristics of the available hardware, and an \emph{implementation model}, defining the relation between the available hardware and the application model.

\begin{wrapfigure}{R}{4cm}
	\centering
	\vspace{-0.7cm}
	\includegraphics[width=3.5cm]{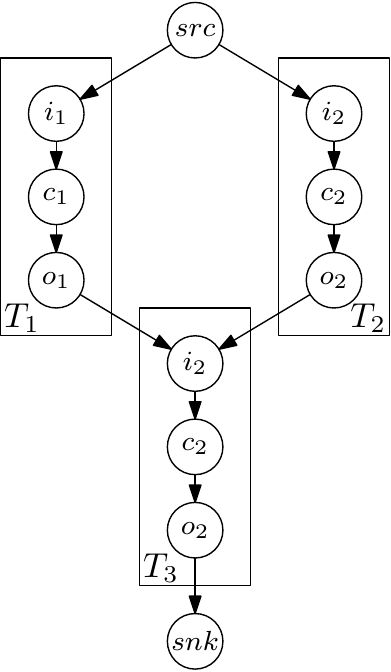}
	\captionof{figure}{Sample memory-augmented task graph with three tasks, one source and one sink.}
	\label{fig:applicationmodel}
\end{wrapfigure}

The \textbf{application model} is based on a task graph, i.e., a directed acyclic graph, where nodes represent tasks and edges represent data dependencies between these tasks. Similarly to Campeanu et al.~\cite{campeanu2014}, we differentiate between \emph{computation nodes} and \emph{memory nodes}. While computation nodes indicate that a certain computation must be executed, memory nodes indicate that data must be made available. More precisely, each tasks consists of three nodes: a memory node representing the input data, a computation node, and a memory node representing the output data. Furthermore, additional memory nodes may indicate data sources or sinks (Figure~\ref{fig:applicationmodel}). This representation is based on the assumption that a high amount of data needs to be computed, making memory access mandatory during the execution of each task. It allows us to accurately differentiate between the cost caused by the computation and the cost caused by the memory access. In particular, it allows us to consider different locations for the data. For example, a CPU could work on data provided by the main RAM and write it back directly into the GPU RAM.

In the \textbf{hardware model}, we assume that (1) each computation device is connected to (at least) one associated memory, (2) data transfer can only happen between memories (not between computation devices) and (3) the computation of a device is blocked by a memory transfer from or to the associated memory.
Usually, the associated memory refers to a respective RAM unit, e.g. a GPU RAM for the GPU or the System RAM for the CPU. The model, however, is not limited to one memory unit per device.
While the data transfer between different memories is usually done through DMA units, it is still reasonable to assume that computation units are affected by the memory transfer, since they cannot access their respective data. Excess data rate, however, can be used to start independent tasks. We elaborate on this in Section~\ref{ssc:databusses}.

The \textbf{(task) implementation model} represents the relation between the application and hardware model. It mainly consists of functions setting elements of both models into relation. Its main purpose is to work as an interface between those two models and to make parts of the modeling framework more interchangeable.
Between each node of the application model and each device, a compatibility relation is defined that indicates which task can be mapped onto which device. Naturally, memory nodes can only be mapped onto memories and computation nodes must be mapped onto a processing device. However, there can be further restrictions. For example a cache may only fit memory nodes that contain a small amount of data or a tensor processing unit can only execute a small subset of tasks.
In addition to a compatibility function, the implementation model defines how much time is needed to execute a task on a certain device. 

The overall advantage of the described modeling approach lies in the possibility to easily evaluate a given task mapping while abstracting from implementation and platform details. Consequently, we define a \textbf{cost function} based on a simple, but reasonably effective, evaluation algorithm.
\begin{algorithm}
\setstretch{1.1}
\SetKwInOut{Input}{Input}
\SetKwFunction{SNodes}{SortedNodes}
\SetKwFunction{Nodes}{Nodes}
\SetKwFunction{Time}{t}
\SetKwFunction{Devices}{Devices}
\SetKwFunction{Mapping}{Map}
\SetKwFunction{Topsort}{topsort\_bfs}
\SetKwFunction{Succ}{successor}
\SetKwFunction{Succs}{successors}
\SetKwFunction{TimeFn}{time}

\Input{\Nodes, \Devices, $\Mapping: \Nodes \mapsto\Devices$}
\SNodes $\leftarrow$ \Topsort{\Nodes}

\lForEach{$p\in\Devices$}{$\TimeFn(p)\leftarrow 0$}
\ForEach{$i\in\SNodes$}{
	\uIf{$i$ \textup{\textbf{is}} input memory node}{
		$j\leftarrow\Succ(i), k\leftarrow\Succ(j)$\\
		$p_i\leftarrow\Mapping(i), p_j\leftarrow\Mapping(j), p_k\leftarrow\Mapping(k)$\\
		$\Time\leftarrow\max(\TimeFn(p_i),\TimeFn(p_j),\TimeFn(p_k)) + d_{i,p_i,p_j} + t_{j,p_j} + d_{j,p_j,p_k}$\\
		$\TimeFn(p_i), \TimeFn(p_j), \TimeFn(p_k)\leftarrow \Time$
	}
	\ElseIf{$i$ \textup{\textbf{is}} output memory node \textup{\textbf{or}} source}{
		\ForEach{$j \in \Succs(i)$}{
			$p_i\leftarrow\Mapping(i), p_j\leftarrow\Mapping(j)$\\
			$\TimeFn(p_i), \TimeFn(p_j)\leftarrow \max(\TimeFn(p_i),\TimeFn(p_j)) + d_{i,p_i,p_j}$
		}
	}
}
\KwRet{$\max_p(\TimeFn(p))$}

\caption{Computation of the total cost of a given task-device mapping.}
\label{alg:simulation}
\end{algorithm}
Algorithm~\ref{alg:simulation} shows the computation of the cost of a given mapping. For each device, a decoupled time value is managed, which is increased when the device is in use. Tasks are queued for execution according to a topological sorting based on a breadth-first search. There are two main cost factors. The \emph{transportation of data} from task $i$ on device $p$ to device $q$, denoted by $d_{ipq}$, and the \emph{execution of a task} $i$ on device $p$, denoted by $t_{ip}$.
Transportation of data happens along the edge between two memory nodes. The time values of both memories are synchronized and increased according to the time given by the implementation model. The time for the execution of a task consists of the time for the read access to the input memory, the write access to the output memory and the computation time on the given device. The time values of all three involved devices are synchronized and the total time for the execution is added to each of them. Note that the input memory waits for the output memory and vice versa to account for the fact that data is processed in small chunks.

After all tasks have finished, the overall cost for the computation is given as the maximum time value over all devices. This value may depend on the used schedule, i.e. the order of tasks in the topological sorting. A potential bias can be circumvented by choosing the topological sorting at random. 

\subsection{Models for different design stages}
\label{ssc:designstages}

The high abstraction level of the model presented in Section~\ref{ssc:abstractmodel} allows the designer to reuse optimization algorithms written for this model in different design stages. In an early design stage, the time for task execution and data transport can be determined based on superficial characteristics of the given tasks and potential devices. This allows for a rapid estimate on the required characteristics for a performance gain and, in consequence, supports the designer in their hardware choice.
In a later design stage, promising tasks may be implemented and measured on different devices. With these more precise values, the same algorithms can support the designer in finding the optimal configuration.

We present a simple realization of the abstract sytem model that can be used during an \textbf{early design stage}. In particular, we describe a more detailed hardware and application model that fulfills the specifications demanded by the abstraction.
The model is primarily based on the \emph{task sizes} of the given application and the processable \emph{data rates} of the devices. The general idea is to get an estimate of the processing time of a certain amount of data based on device characteristics.
Each task node is attributed with a \emph{data processing function}, which computes the amount of output data generated from input data of a certain size, e.g. a simple sum of two values would have a 2:1 relation between input and output data. In addition, each node has a \emph{complexity function}, which determines the amount of computations needed based on the input data.
Finally, each computation node indicates which percentage of its execution time is \emph{parallelizable}. For the sake of simplicity, we assume that the parallelizable part is fully parallelizable with an arbitrary amount of processors.

In the hardware model, we compute the data rate of a memory as the product of (1) the bus clock speed, (2) the bus width and (3) the number of memory channels. We set the \emph{serial data rate} $\serialrate$ of a processing device to the clock rate multiplied with a device-specific \emph{overhead penalty}, describing the overhead caused by the microarchitecture. Note that a penalty is relevant only if the overhead is expected to be vastly different between devices. In the evaluation given in Section~\ref{sec:evaluation}, we therefore do not apply penalties.
In addition to the serial data rate, each processing device is assigned a \emph{parallelization factor} $\parallelfactor$ consisting of (1) the number of cores and (2) the potential data parallelism. For example, in case of a GPU, the second factor equals the number and width of SIMD units.

Finally, in the implementation model, we set the \emph{execution time} of a task node on a device to $0$ for a memory node and to $\data_{in} / (\serialrate * (1-\parallelizability + \parallelizability\parallelfactor))$ for a computation node, where $p\in[0,1]$ denotes the parallelizability of the task. The \emph{transport time} is determined by the minimum of the data rates of the two connected devices and a potential data rate limitation between them. It is set to infinity if no edge is present in a given hardware graph.

Using this model, an early assessment of the potential of a heterogeneous implementation can be made. In a \textbf{later design stage} a measure-based model should replace these rough estimates. For this, (time) complexity functions for both the execution and the transport time should be derived from the measured data, which can then be directly incorporated into the task implementation model. Using appropriate penalties, a mixture of both models can be used if measured data isn't available for all task-device combinations.

\subsection{Extension: Full usage of data busses}
\label{ssc:databusses}

Data transport between two memories is usually done through DMAs, which are independent of the processing devices. Hence, processing devices are in principle able to execute tasks during the transport of (independent) data. In the presented abstract model, on the other hand, we wait until the input and output memories are unoccupied before we start another execution. The reasoning behind this decision is that during processing, data must be accessed by the processing device and therefore access to the memory bus is needed. However, a data transaction does not always use the full data rate of both memories. If, for example, memory is transferred between System RAM and GPU RAM, the transaction speed is usually limited by the bus of the GPU RAM. The remaining bus width of the System RAM can be used by a processing device to access data.

The resulting gain in performance can be incorporated into the model by adjusting the blocking time according to the used ressources. Let $r_1, r_2$ be the data rate of two devices $p_1,p_2$ with $r_1\leq r_2$. Then a data transport between these two devices that takes time $t$ increases (after synchronization) the time value of $p_1$ by $t$ and of $p_2$ by $\frac{r_1}{r_2}t$.
The increase in the time value of $p_2$ represents the time that the device would work if it could use all of its ressources for the task, i.e., the total delay that a parallel execution of other tasks accessing $p_2$ would experience.
Note that the additional capabilities can only be used by independent computations. A task that is dependent on the data transport between $p_1$ and $p_2$ won't be able to make use of the free resources. Hence, the cost computation algorithm must assure that a dependent task waits the full time $t$ until its computation is started.

\subsection{Extension: Streamability and Virtual Memory}

In the current model we write data back to the memory after each task execution. Depending on the granularity of the tasks, this may be inefficient if a subsequent task is executed on the same device. If a task works only locally on the given data, we may do several subsequent processing steps on the same data before writing it back to memory. These tasks are called \emph{streamable}. We can model this behavior in two ways: (1) we modify the cost function to ignore memory accesses between subsequent tasks that are executed on the same device and do not produce intermediate data used by other devices or non-streamable tasks, or (2) we introduce \emph{virtual memories} into the hardware model with zero access time from the chosen device and infinite data transfer time to other devices. Virtual memories can then be used in between operations on the same device to hide the memory access.
The first variant increases the complexity of the cost function, whereas the second variant shifts the responsibility to the mapping algorithm.

A special case for streamability is the handling of dataflow-based devices such as FPGAs. Here, not only the memory access can be omitted, but also the execution of tasks can be pipelined, i.e., operations can be executed in parallel along the stream. Therefore, a subtree of streamable tasks on such a device will only take as long as the most expensive processing or memory node in the subtree. A limitation to this property is given by the limited area on such a device. To integrate this behavior into our model, we introduce an area requirement for all tasks and modify the cost function to compress subtrees up to the size of the respective device to one single task.
Furthermore, bigger tasks that are streamable and fit on a single FPGA may also greatly benefit from pipelined processing. Regarding Section~\ref{ssc:designstages}, the behavior can be modeled by a streamability factor for each task, indicating into how many pipelined steps the task can be split. If a computation node is mapped onto an FPGA and doesn't exceed the maximum area available on the FPGA, the execution time is reduced by this factor.

\section{Mixed-Integer Linear Programs}

The abstract model presented in Section~\ref{ssc:abstractmodel} allows us to effectively develop and compare algorithms and heuristics for heterogeneous task assignments without regard for implementation details. In this section, we present two mixed-integer linear programs for heterogeneous task assignment based on the model.

\subsection{Device-based ILP}

In the first MILP, we aim to minimize the maximum time on each device. In a system with $n$ nodes and $m$ devices, let $t_{ip}$ be the time required to execute task $i$ on device $p$ and let $d_{ipq}$ be the time required to transport the output data of task $i$ from device $p$ to device $q$. 
Let $x_{ip}$ be a binary variable indicating that task $i$ is executed on device $p$, and let $E$ be the set of edges in the application graph. 
Then the times $T_p, T_{p}^{in}, T_{p}^{out}$ reflecting the total time of execution on, transport to, and transport from device $p$, respectively, are given as:
\begin{align*}
T_p = \sum_{i=1}^n x_{ip}t_{ip} \qquad T_{p}^{in} = \sum_{q=1}^m \sum_{(i,j)\in E}d_{ipq}x_{ip}x_{jq} \qquad T_{p}^{out} = \sum_{q=1}^m \sum_{(i,j)\in E}d_{iqp}x_{iq}x_{jp}
\end{align*}

Note that the quadratic terms $x_{ip}x_{jq}$ can be replaced by single variables using the McCormick inequalities $x_{ipjq}\leq x_{ip}$, $x_{ipjq}\leq x_{jq}$ and $x_{ipjq}+1\geq x_{ip}+x_{jq}$. 
Our goal is to minimize the term $\max_p(T_p+T_{p}^{in}+ T_{p}^{out})$. To resolve the minmax formulation, we introduce another variable $z$ with $z\geq T_p+T_{p}^{in}+ T_{p}^{out}$ for all $p\in\lbrace 1,...,m\rbrace$, which is then minimized.
As additional constraint we ensure that each task node is mapped to one device. Let $C_i$ be the set of devices that are compatible to task $i$. Then we want to guarantee that $\sum_{p\in C_i} x_{ip} = 1$ for all $i\in\lbrace 1,...,n\rbrace$. Hence our final MILP is given as
\begin{align*}
\textrm{\textbf{minimize} } &z \\
\textrm{\textbf{subject to} } &z\geq T_p+T_{p}^{in}+ T_{p}^{out} &\forall p\in\lbrace 1,...,m\rbrace \\
&\sum_{p\in C_i} x_{ip} = 1 &\forall i\in\lbrace 1,...,n\rbrace
\end{align*}

\subsection{Time-based ILP}

While above MILP is reasonably simple, it does not consider execution order and synchronization issues. In this section, we present a more exact, but also more expensive time-based linear program. Here, the goal is to \enquote{simulate} an execution, i.e., to assign start and end times to each task. For this, we introduce variables $y_{i,0}, y_{i,1}$ representing the start and end of the execution of node $i$. 

With the notation from the previous section, we guarantee that there is sufficient time before the start and the end of the execution of a node and that a node can only be started if all previous nodes have been processed. Hence, 
\begin{align*}
	y_{i,1} \geq y_{i,0}+\sum_{p=1}^m x_{ip}t_{ip} \quad \textrm{ and } \quad y_{j,0} \geq y_{i,1}+\sum_{p=1}^m\sum_{q=1}^m d_{ipq}x_{ip}x_{jq}
\end{align*}
for all tasks $i$ and all edges $(i,j)$, respectively. In contrast to the device-based variant, we must assure that each device is used for only one task simultaneously. For this, we sort the tasks topologically and assure that all tasks that are mapped onto the same device are executed in topological order. Hence, we demand
\[
y_{j,0} \geq \sum_{p=1}^m x_{ip}x_{jp}y_{i,1} 
\]
for all $j\in\lbrace 1,...,n\rbrace$ and all $i < j$. This equation can be linearized by replacing it with $y_{j,0}-y_{i,1} \geq Mx_{ip}x_{jp}-M$ for all $p$ with a sufficiently large constant $M$ and using the McCormick inequalities as before.
We minimize the maximum time $z$ by demanding $z\geq y_{i,1}$ for all tasks. Adding, as before, the condition that a device must be assigned to each task node, we get
\begin{align*}
\textrm{\textbf{minimize} } &z \\
\textrm{\textbf{subject to} } &z\geq y_{i,1} &\forall i\in\lbrace 1,...,n\rbrace \\
&y_{i,1} \geq y_{i,0}+\sum_{p=1}^m x_{ip}t_{ip} &\forall i\in\lbrace 1,...,n\rbrace \\
&y_{j,0} \geq y_{i,1}+\sum_{p=1}^m\sum_{q=1}^m d_{ipq}x_{ip}x_{jq} &\forall (i,j)\in E \\
&y_{j,0} \geq \sum_{p=1}^m x_{ip}x_{jp}y_{i,1} &\forall j\in\lbrace 1,...,n\rbrace, \forall i\in\lbrace 1,...,j-1\rbrace\\
&\sum_{p\in C_i} x_{ip} = 1 &\forall i\in\lbrace 1,...,n\rbrace
\end{align*}

\subsection{Extension: Streamable devices}

The time-based linear program can be extended to reflect the pipelining behavior of dataflow-based devices such as FPGAs. For this, we modify the order constraint to enable tasks on streamable devices to start simultaneously with a parent task executed on the same device. Let $D$ be the set of all devices and $D_p$ be the set of pairs of dataflow-based devices and their associated memories (including pairs with themselves). Then the modified constraint is given as
\begin{align*}
y_{j,0} \geq y_{i,1}+\sum_{(p,q)\in D^2\backslash D_p} d_{ipq}x_{ip}x_{jq} - \sum_{(p,q)\in D_p}x_{ip}x_{jq}t_{ip}, \quad \forall (i,j)\in E
\end{align*}
By this, we effectively reduce the constraint to $y_{j,0} \geq y_{i,0}$ if both tasks are on associated dataflow-based devices. Finally, to take the maximum of all operations in the pipeline, we ensure that a task cannot end before its parent ends, i.e., $y_{j,1} \geq y_{i,1}\; \forall (i,j)\in E$.

Since devices such as FPGAs have a maximum capacity, we must ensure that the total number of tasks added to the device does not exceed this capacity. Let $s_i$ be the area requirement for task $i$ and $S_p$ be the capacity of a device $p$. Then 
\[
\sum_{i=1}^n x_{ip}s_{i} \leq S_p \quad \forall p\in D_s
\]
where $D_s$ is the set of all streamable devices. This capacity constraint is added to the device-based approach as well to ensure a valid configuration, even though the pipelining capability can't be represented.

\section{Evaluation}
\label{sec:evaluation}

We demonstrate the usage of the model in an early design stage through several experiments in a sample environment. We determine the execution time and data transfer time based on the specifications of the given devices and the size of a virtual data load as described in Section~\ref{ssc:designstages}. Our virtual test system contains an AMD Epyc 7531P with 16 cores (32 threads), a clock rate of \SI{2.4}{\giga\hertz} and SIMD processing with $\num{8}\times\SI{32}{\byte}$ words, as well as a AMD Radeon RX Vega 56 with \SI{1.6}{\giga\hertz} and \num{3584} SIMD units. Furthermore we assume a Xilinx XCZ7045 FPGA with a clock rate of \SI{400}{\mega\hertz} and an equivalent of \SI{350}{\kilo\nothing} logic cells, partitioned into \num{28} area units. 
We assume appropriate RAM units for CPU, GPU and FPGA with a calculated throughput of \SI{170}{\giga\byte\per\second}, \SI{410}{\giga\byte\per\second} and \SI{11}{\giga\byte\per\second}, respectively.

For the application, we generate random series-parallel graphs with \num{30} edges. For this, we start with a connected source and sink node and subsequently add edges using either a series (split an edge into two by adding a node on it) or parallel (copy an edge) operation. The resulting graphs are stereotypical for data-intensive applications where you start with a common data set, process the data along different computation paths and combine the outputs to a common result. In order to avoid duplicate edges, we set the probability of a series operation to $0.5 + 0.5\frac{i}{m}$, where $m$ is the desired number of edges and $i$ is the number of edges already added. That is, we start with a probability of \num{0.5} and continuously increase the probability to \num{1}. After removing duplicate edges, we arrive at graphs with, on average, around \num{21} nodes and \num{22} edges. Each node, except for the source and sink, is then converted to a task with input, computation and output node, resulting in application graphs with on average slightly below \num{60} nodes.

We assign the same data load to each task, so the data processing function of each task is the identity function. We choose the parallelizability of a task uniformly between \num{0} and \num{1} and the complexity function as a linear function $f(x)=cx$, where the factor $c$ is log-normal-distributed with $\mu = 3, \sigma = 0.5$. The parameters are chosen to create generally similar complexities with occasional outliers of significantly higher complexity. About \SI{90}{\percent} of the generated values for $c$ lie in the interval $[10,50]$ with a median of $~20$. For the FPGA extension, we assume that every task is streamable and that the area needed for a task as well as the possible gain through streaming is equal to its complexity factor. Through this, one used unit of area is equated to roughly one pipelining step.

\begin{table}[htb]
\centering
\newcolumntype{Y}{>{\centering\arraybackslash}X}
\vspace{-0.5cm}
\begin{tabularx}{0.85\textwidth}{l|YYYY|Y}
					& avg.				  & min.                & max.				 & \# impr.	& time avg. \\
\hline\hline
\textbf{CG} &&&&\\
Device-based        & \SI{-10}{\percent}   & \SI{-19}{\percent}  & \SI{29}{\percent}   & 8      & \SI{0.06}{\second}   \\
Time-based          & \SI{11}{\percent}    & \SI{-10}{\percent}  & \SI{66}{\percent}   & 81     & \SI{4.98}{\second}   \\
\hline\hline
\textbf{CGF} &&&&\\
Device-based        & \SI{1}{\percent}     & \SI{-17}{\percent}  & \SI{54}{\percent}   & 46     & \SI{0.11}{\second}   \\
Time-based          & \SI{17}{\percent}    & \SI{-8}{\percent}   & \SI{64}{\percent}   & 92     & \SI{18.95}{\second}   \\
\hline\hline
\textbf{CGFF} &&&&\\
Device-based        & \SI{6}{\percent}     & \SI{-18}{\percent}  & \SI{40}{\percent}   & 65     & \SI{0.19}{\second}   \\
Time-based          & \SI{19}{\percent}    & \SI{-5}{\percent}   & \SI{71}{\percent}   & 94     & \SI{47.54}{\second}  \\
\end{tabularx}

\caption{Performance gain through task assignment strategies compared to assigning all tasks to the CPU for \num{100} graphs with on average {\raise.17ex\hbox{$\scriptstyle\mathtt{\sim}$}}\num{20} tasks and \SI{100}{\mega\byte} input data. The fourth column indicates the number of cases in which the performance could be improved. Execution time is given in the last column.}
\vspace{-1cm}
\label{tbl:results}

\end{table}

In Table~\ref{tbl:results}, results are listed for three different hardware configurations: A configuration with only CPU and GPU (\textbf{CG}), a configuration with CPU, GPU and one FPGA  (\textbf{CGF}) and a configuration with CPU, GPU and two identical FPGAs (\textbf{CGFF}). It shows the average, minimum and maximum change of performance compared to an implementation where all tasks are mapped to the CPU. For our input data, mapping all tasks to the GPU makes the execution about \SI{33}{\percent} slower. Compared to the CPU, the higher parallelization factor of the GPU leads to an improvement only if close to \SI{100}{\percent} of the task is parallelizable. Consequently, potential improvements through the GPU are mainly enabled by the simultaneous execution of different tasks using uncontended memories.

As the results show, the time-based ILP is usually more effective than the device-based ILP in increasing the performance of the execution. Both the maximum performance gain and the frequency of getting an improved mapping is higher for the time-based ILP. Adding one or two FPGAs increases the size of the design space and consequently leads to more optimization opportunities, showing potential performance gains of up to \SI{71}{\percent}.
\begin{figure}[htb] 
	\centering
	\includegraphics[width = 0.45\textwidth, trim={1.3cm 1.3cm 1.3cm 1.3cm},clip]{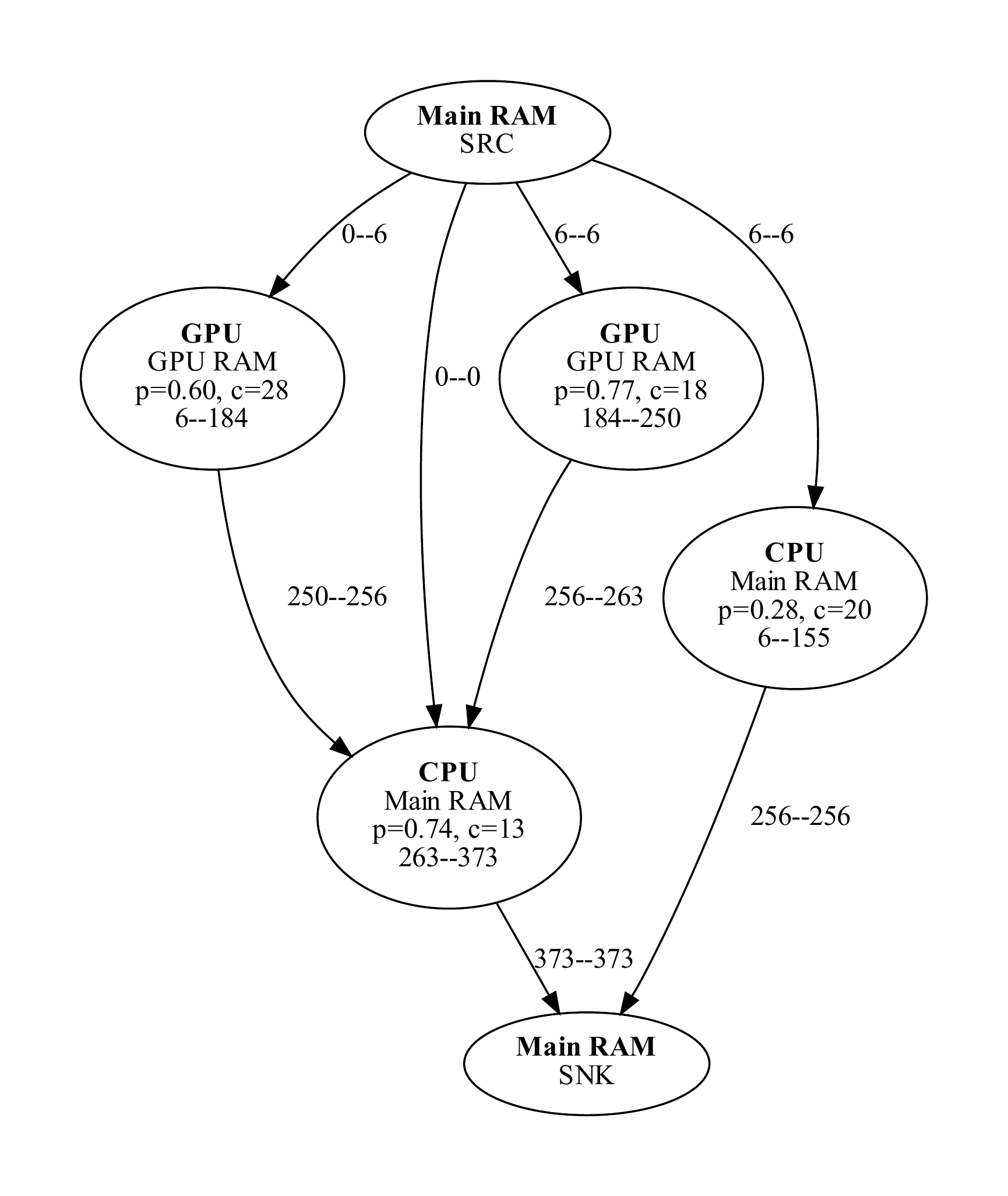}
	\includegraphics[width = 0.45\textwidth, trim={1.3cm 1.3cm 1.3cm 1.3cm},clip]{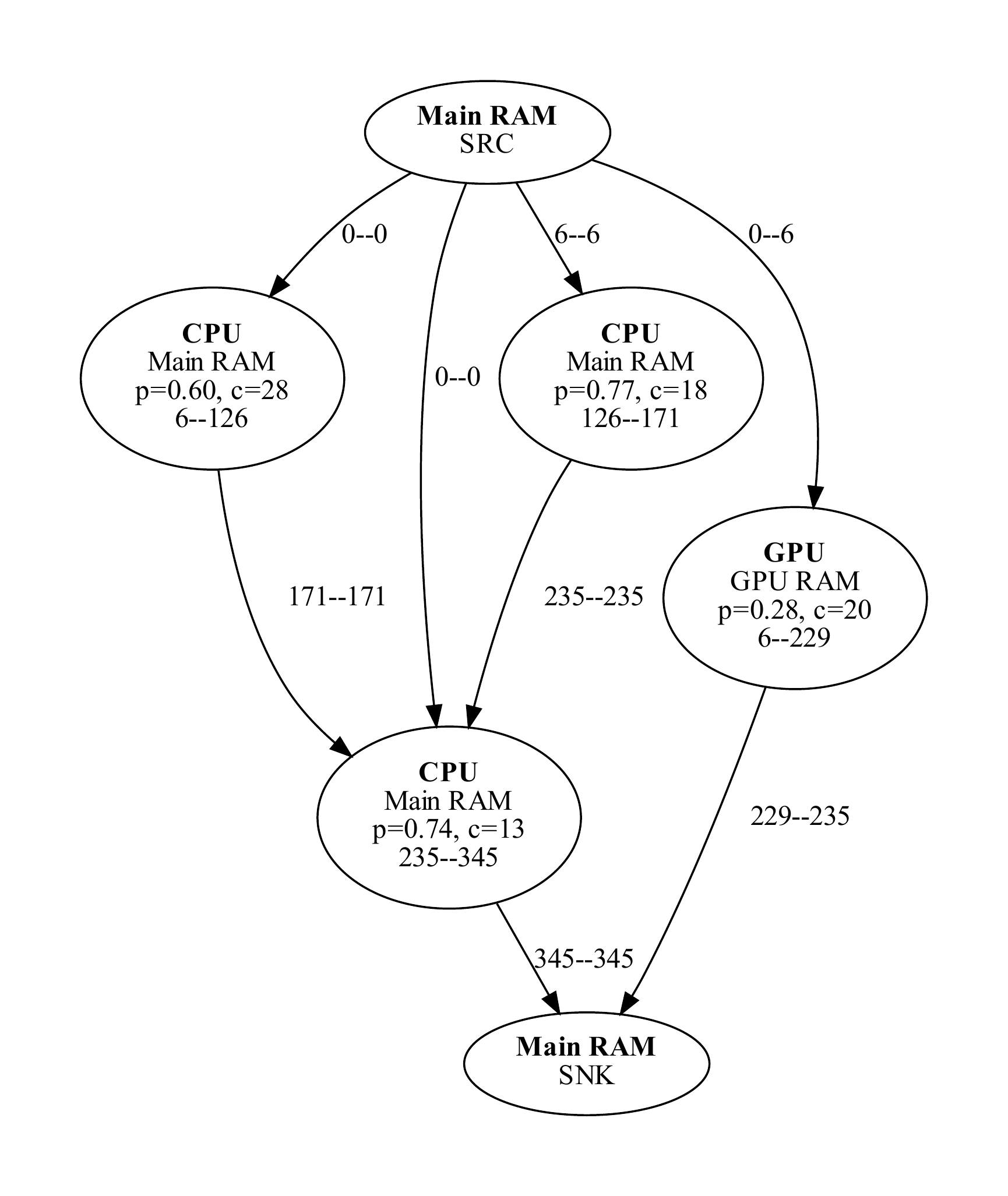}
	\caption{Mappings found by the \textbf{device-based} (left) and the \textbf{time-based}  (right) LP for a small sample graph. For each node, the parallelizability $p$ and complexity factor $c$ are given. At the edges and in the nodes, the time windows for transport and computation are annotated. For all tasks, the chosen input and output memory are identical, the corresponding nodes are omitted for readability.}
	\vspace{-0.5cm}
	\label{fig:mappingexample}
\end{figure}
An exemplary mapping of the two algorithms is shown in Figure~\ref{fig:mappingexample}. Both depicted mappings improve on a pure CPU mapping (which has a cost of \num{423} time steps). The device-based approach chooses to put two moderately well parallelizable tasks on the GPU (with parallelizability \num{0.6} and \num{0.77}, respectively). However it is not able to recognize that both nodes lie on the critical path of the task graph. The time-based approach is able to identify the critical path and therefore puts a badly parallelizable, but uncriticial, task on the GPU, reducing the overall cost of the mapping. However, there are cases in which the device-based ILP finds a better mapping, since it is not restricted to follow a specific topological order. Furthermore, it is less complex to solve and therefore better suitable for very large task graphs.
The linear programs are solved using Gurobi~9.1.2~\cite{gurobi} in Python on an AMD EPYC 7542 with \SI{2}{\tera\byte} RAM. The execution time increases quickly with the size of the task graph. As shown in Table~\ref{tbl:results}, the device-based approach is about two orders of magnitude faster than the time-based approach.

In the example shown in Figure~\ref{fig:mappingexample}, the transfer cost between different memories has only a small impact on the mapping. This changes drastically if the complexity of the computations is reduced. If the complexity is set to \num{1} for all tasks, switching devices is much more costly compared to the computation itself. In this case, in each of the hardware configurations only about \num{40} out of \num{100} graphs with \num{30} edges could be improved using the time-based algorithm and about \num{2} out of \num{100} graphs with the device-based ILP. Furthermore, the tendency to map multiple connected tasks to the same device strongly increases.

\section{Conclusion}

The model presented in this work provides a solid basis for the development of general task assignment algorithms. A common model allows the designer to use various heuristics to explore the design space for potential improvements early in the design process. In particular, a large database of available algorithms helps in deciding early on whether a potential optimization is worth the effort.
The realization of the model in different design stages currently still puts much responsibility to the designer. The modeling of the time function assessed in Section~\ref{ssc:designstages} provides a direction on how the model can be used. The development of more precise realizations is open for future research.
The given MILPs are sufficiently powerful to find significant improvements for small task graphs. Furthermore, they form a robust baseline to assess the quality of future heuristics for large task graphs or dynamic ressource allocation in a changing environment.

\bibliographystyle{splncs04}
\bibliography{literature}

\end{document}